\documentstyle[12pt,epsfig,graphicx,citesort]{cernart}
\begin{document}
\begin{titlepage}

\vskip 2.3cm
\begin{flushright}
  \parbox{0.36\hsize}{
    \large \bf 
    ALICE--INT--2003--37\\
    Internal Note/PHOS\hspace*{-2.1cm}\\
    27 October 2004
    \\
  }
\end{flushright}

\vskip 1.5cm

\title{\Large { \bf Direct Photon Identification with  
           Artificial Neural Network in the Photon Spectrometer PHOS}}
\begin{Authlist}
\large{
  M.Yu.Bogolyubsky,
  Yu.V.Kharlov,
  S.A.Sadovsky
  
  \vspace{0.2cm}  
  Institute for High Energy Physics,
  Pobeda str., 1, \\
  Protvino, 
  142281, Russia
}
\end{Authlist}

\vskip 1.5cm

\begin{abstract}
  A neural network method is developed to discriminate direct photons
  from the neutral pion background in the PHOS spectrometer of the
  ALICE experiment at the LHC collider.  The neural net has been
  trained to distinguish different classes of events by analyzing the
  energy-profile tensor of a cluster in its eigen vector coordinate
  system.  Monte-Carlo simulations show that this method diminishes by
  an order of magnitude the probability of $\pi^0$-meson
  misidentification as a photon with respect to the direct photon
  identification efficiency in the energy range up to 120~GeV.
\end{abstract}
\end{titlepage}
 
\section{Introduction}

The Large Ion Collider Experiment ALICE \cite{ALICE} is intended to
study heavy-ion collisions at the energy of 5.5 TeV per nucleon at the
LHC collider.  For photons detection ALICE is equipped by the PHOton
Spectrometer (PHOS) \cite{PHOS-TDR} which is a high-granularity
electromagnetic calorimeter built of lead-tungstate crystals
(PbWO$_4$) with the transverse dimension of $2.2\times 2.2$ cm$^2$ and
the length of 18~cm.  The PHOS detector consists of five modules, each
made as a matrix of $56\times64$ crystals located at 460~cm from the
beams interaction point. The spectrometer is positioned at the bottom
of the ALICE set-up covering $|\eta|<0.135$ in pseudorapidity and
100$^\circ$ in the azimuthal angle.

The ALICE set-up is rather transparent to $\gamma$-radiation. To reach
PHOS, the produced photons pass through the Inner Tracking System
(ITS) \cite{ITS}, the Time Projection Chamber (TPC) \cite{TPC} and the
Charge Particle Veto Detector (CPV) \cite{CPV}.  The stratum of medium
takes $\sim 0.1$ of the radiation length \cite{AN-2001-37} (with the
main contribution from ITS and TPC), which gives a small probability
of the secondary photon production in the medium.

One of the tasks of the ALICE experiment is to detect direct photons
carrying the information about fundamental processes at extreme
conditions of the quark matter.  The energy range of such photons
extends from 0.4 GeV to 10 GeV for thermal radiation of Quark Gluon
Plasma (QGP) and higher, up to hundreds GeV, for radiation occurring
also in early collisions of the constituents.  In the latter case the
essential background is arising from the two-photon decay of
$\pi^0$-mesons, produced at the same energies as photons, due to the
merge of decay photons into one shower in PHOS.  Contribution of the
photonic decay from heavier meson resonances ($\rho$, $\eta$,
$\omega$, etc. ) exists as well.  Such photons provide a rather heavy
background to direct photon production over the whole $p_T$ range and
they should be subtracted from the data.  It leads to the necessity of
identifying the PHOS showers as those produced by the photons or
$\pi^0$-mesons.

The straightforward way to discriminate the considered background from
direct photons is to exploit a powerful reconstruction program tuned
especially for the given PHOS structure and able to resolve the fine
problem of the "direct" photon selection (say, by the maximum
likelihood method), and correspondingly, to find the real number of
the showers even in the overlapped clusters.  To accomplish this task,
a program should correctly take into account the shower profiles of
photons and charged particles at different incident angles, as well as
their fluctuations, the electronics noise, the threshold on registered
signals (that increases the number of zero channels in the selected
cluster), etc. In practice such a perfect program is often too
complicated for fast realization with long tuning and commissioning
and additionally needs essential computational resources.  Therefore,
real reconstruction programs are usually created in some
simplification assumptions which results in the degradation of its
discrimination power for direct photon selection.  Such simplified
programs are often based on the recognition of the shapes of the
showers produced by different kinds of particles.
 
In this paper we apply the Artificial Neural-Network (ANN) approach
\cite{NN} for the direct photon identification in PHOS. The ANN-method
has already recommended itself as a powerful tool in different
applications of high energy physics, e.g.
%YVK2: nobody has separated quarks from gluons yet.
%YVK2: maybe one meant ``separation of jets produced by quarks and gluons''?
quark and gluon jet separation,
%YVK2: same for b-quark: precise here
b-quark identification, Higgs boson search, selection of the rare
neutral meson decay modes violating $C$-parity, etc.
\cite{1,2,3,4,5,6}.
 
The main peculiarity of our method is the use of the energy-profile
tensor of a cluster, which components are calculated in its
eigen-vector coordinate system, with the aims of the neural net
training and the afterward event classification. Our calculations show
essential recognition capacities of this procedure that were examined
using a sample of Monte-Carlo generated events simulating the
isolating production of direct photons and $\pi^0$-mesons for the real
ALICE set-up.

\section{Application of neural network algorithm method}

In the analysis of experimental data, a standard procedure of
selecting signal events is based on various cuts of observed
kinematics variables.  A general case of such cuts corresponds to a
particular set of functions, called feature functions or neurons.  In
general terms, neural networks are represented as a large number of
interconnected functional units named nodes with a neuron in each of
them. The data processing is organized in the most common, to-date,
architecture called the Multilayer Perception (MLP).  MLP incorporates
one input layer through which the initial data (features) are
injected, several hidden layers, and one output layer of neurons.

Output response $O_i$ of the $i$-th neuron is obtained by the so
called sigmoid function $f$ $(0\le f \le 1 )$ dependent on the
weighted sum of all input signals $S_j$ to this neuron:
\begin{equation}
  O_i=f(\sum\limits_j w_{ij}\cdot S_j +\theta_i)~,
  \label{Oi}
\end{equation}
where index $j$ runs over all the inputs of the $i$-th node, 
$w_{ij}$ and $\theta_i$ are the weights and the correspondent thresholds 
optimizing the selecting power of the procedure. Function $f$ is
defined as 
\begin{equation}
  f(x)=(1+\tanh(x))/2~.
  \label{f}
\end{equation}
The output layer, consisting of one node, provides a neural-net
response $S_{NN}$, wedged between 0 and 1, and used to classify the
events.

The use of the neural network is a two-step process, i.e. a learning
stage followed by an application stage. During the learning phase
using the Monte-Carlo simulation, we know about every event whether it
is a background or a signal one. The optimal values of weights
$w_{ij}$ and thresholds $\theta_i$, (see (\ref{Oi})), for resolving
the problem are determined by minimizing the functional
$L(w_{ij},\theta_i)$
\begin{equation}  
  L(w_{ij},\theta_i)=\frac{1}{2}\sum\limits_{k=1}^N 
  |S_{\rm learn}^{(k)}-S_{NN}^{(k)}|^2~,
  \label{L}
\end{equation}
where index $k$ runs over all $N$ training events, $S_{\rm
  learn}^{(k)}=0$ for the background and $S_{\rm learn}^{(k)}=1$ for
desired events, $S_{NN}^{(k)}$ are the ANN response. The details of
the whole minimization procedure can be found in \cite{7}. And finally
the quality of learning is tested with an independent sample of
Monte-Carlo events.

\subsection{Input variables for the ANN}

Reconstruction programs for cellular electromagnetic calorimeters
provide data as a set of clusters defined as a group of cells with a
common edge. Every cluster is characterized by the amplitudes of
signals from the cells and coordinates of the cluster cells on the
detector plane.  The total length of the data array for one cluster is
$3N_{c}$, where $N_{c}$ is a number of cells in the cluster. This
array contains exhausting experimental information about the cluster
but, however, it is not convenient for the aim of the ANN approach, at
least due to the varying data length from event to event. To overcome
the latter problem one can use only the limited number of major
cluster cells (say, $3 \times 3$ around the cell with the maximal
amplitude \cite{CMS_NN} ), but it definitely leads to the uncontrolled
information loss.

The essential part of our ANN-approach to $\gamma/\pi^0$ selection is
to choose such a fixed number of variables that carry, nevertheless,
the principal part of information with its volume big enough to find a
solution and which length is independent of the cluster size.  First
of all each cluster is characterized by a position of its
center-of-gravity $\vec{X}_g$ as well as by the center of the cell
with the maximal signal in the cluster $\vec{X}_m$.  Vector
$\vec{X}_g$ is defined according to the standard formula
\begin{equation}
  \vec{X}_g=\sum\limits_k \vec{x}^{(k)} E_k / \sum\limits_k E_k~,
  \label{Xg}
\end{equation}
where index $k$ runs over all cluster cells, $x_1^{(k)}$ and
$x_2^{(k)}$ being the coordinates of the $k$-th cluster cell in an
arbitrary coordinate system.

Then we introduce an energy-profile tensor $Q_{ij}$ of a cluster (that
can be also interpreted as a $2\times 2$ matrix) calculated in the
local cluster coordinate system.  There are two natural possibilities
to set the origin of this coordinate system: either in the center of
gravity $\vec{X}_g$ or in the center of the cell with the maximal
signal in the cluster $\vec{X}_m$.  In our calculations we choose the
latter option.  After that $Q_{ij}$ is defined as
\begin{equation}
  Q_{ij}=\sum\limits_k x_i^{(k)} x_j^{(k)} E_k;\quad i,j=1,2.
  \label{Q}
\end{equation}
Here index $k$ runs over all the cluster cells, $x_1^{(k)}$ and
$x_2^{(k)}$ are coordinates of the $k$-th cluster cell, for
definition, in non-dimensional units of the crystal transverse size,
and $E_k$ is the energy detected in this cell. Tensor (\ref{Q}) is a
quadratic positive-definite form.

The introduced tensor (\ref{Q}) reflects the cluster energy profile
which depends on the number of showers overlapped in one cluster, as
well as on inclination of the photon incidence on the detector.  To
avoid the latter effect or, at any rate, to decrease it we have made
(before calculating matrix $Q_{ij}$) a compression of the cluster
space relative to the origin of the local cluster coordinate system
$\vec{X}_0=(x_{10},x_{20})$ along the vector directed from the
geometrical center of the PHOS module to the point $\vec{X}_0$.  The
compression coefficient has been defined as $c=\cos \Theta$, where
$\Theta$ is the angle between the perpendicular to the PHOS module
surface and the photon propagation direction. The coordinate
transformation due to the compression operation has been made
according to the formula
\begin{equation} 
  \vec{x}\to R(-\phi_0)\cdot E_c(\Theta)\cdot R(\phi_0)\cdot 
  \left (\vec{x}-\vec{X}_0 \right )+\vec{X}_0,
  \label{compression}
\end{equation}
where $\vec{x}=(x_1,x_2)$ is the vector of coordinates of an arbitrary
transfered point, $R$ and $E_c$ are the matrices of rotation and
compression, respectively:
\begin{equation}
  R(\phi_0)=\left(\matrix{
    \cos\phi_0 & \sin\phi_0 \cr 
    -\sin\phi_0 & \cos\phi_0 \cr}
  \right), \quad
  E_c=\left( \matrix{ \cos\Theta & 0\cr
                             0 & 1\cr}                                              
  \right)
  \label{R_Ec}
\end{equation}
and $\phi_0$ is the polar angle of the point $\vec{X}_0$ in the polar
coordinate system with its origin in the geometrical center of the
PHOS module.

One can calculate two eigen values $ \lambda_1$ and $\lambda_2$
(ordered as $ \lambda_1 \ge \lambda_2$) of $Q_{ij}$ and find the
correspondent eigen vectors $\vec{e}_1$ and $\vec{e}_2$ (normalized to
a unit), defining a new coordinate system, where $Q_{ij}$ is reduced
to the diagonal form.  In this new system we also define moments
$M_{mn}$:
\begin{equation}
  M_{mn}=\sum\limits_k (x{'}_1^{(k)})^m\cdot (x{'}_2^{(k)})^n~E_k;
  \quad m,n=0,1,2, \ldots,
  \label{M}
\end{equation}
where index $k$ runs over the cluster cells, $x{'}_1^{(k)}$ and
$x{'}_2^{(k)}$ are the coordinates of the $k$-th cluster cell in the
new coordinate system.  Note that $M_{20}=\lambda_1$,
$M_{02}=\lambda_2$, and $M_{00}=E=\sum_k E_k$ is the total cluster
energy.  Such important magnitudes as distance $d$ between hits of two
glued photons and their effective mass $M_{\gamma\gamma}$ can also be
expressed through $M_{mn}$ (see addendum).

We would also like to point out an essential remark. Initially the
experimental information was carried by signal amplitudes and cell
coordinates, and now it is represented by the introduced moments
$M_{mn}$.  We shall construct from $M_{mn}$ the input vector
$\vec{P}^{(in)}$ of event features for ANN.  One additional angle
variable, not directly expressed through the $M_{mn}$, can be added.
This is angle $\phi$ between the eigen vector $\vec{e}_1$ and vector
$\vec{X}_0$ directed from the geometrical center of the PHOS module to
the cluster center
\begin{equation}
  \phi=\arccos(~\vec{e}_1,~\vec{X}_0 / | \vec{X}_0 | ~).
  \label{delta}
\end{equation}
The use of this angle, together with the coordinates of the cluster
center on the detector plane, significantly improves the quality of
$\gamma/\pi^0$-selection mainly in the peripheral region of the PHOS
modules.

We have found that there are different sets of variables that allow to
construct effective event feature vectors. They include the total
detected cluster energy, the eigen values $ \lambda_1$ and
$\lambda_2$, the moments $M_{mn}$, the estimation of the effective
mass $M_{\gamma\gamma}$ and distance $d$ between $\gamma\gamma$-hits
expressed through $M_{mn}$ (see formulas (\ref{solution1}) and
(\ref{mef_for}) ), and the angle variable $\phi$. Coordinates of the
cluster center $\vec{X}_0$ relative to the center of the PHOS module
can also be added, which increases the selection power of the method.
   
\subsection{Strategy of the $\gamma/\pi^0$-selection}

In this section the algorithm of the $\gamma/\pi^0$-selection is
discussed in detail. First of all we note, that at low energies there
is a background from $\pi^0$ decays producing two separated clusters
in PHOS that can be taken into account by one of the statistical
methods, based on a good resolution of PHOS for
$M_{\gamma\gamma}$-effective masses in this case, which is
demonstrated below in Fig.\ref{mef_fig}b. The mentioned procedures
can, for example, reject the $\pi^0$-contribution by calculating the
masses of $\gamma\gamma$-combinations and by comparing them with the
$\pi^0$-mass.  It is possible to reduce the number of such
combinations at high occupancy of the detector by taking into account
the characteristic angle of the decay cone for photons.  These methods
are beyond the scope of the current paper.  Thus, further calculations
were fulfilled following the assumption that this type of backgrounds
had been eliminated correctly.

Besides, there are two other essential sources of the background due
to $\pi^0$-decays, when only one cluster appears in PHOS.  The first
one results from the detector geometry, i.e. when one of the decay
photons from $\pi^0$ escapes detection in PHOS due to the limited
acceptance, whereas the second photon hits PHOS and generates a single
shower and therefore a single cluster. It provides actually
indistinguishable from the direct photons sample of the background
photons in the whole photon energy range.  This background is
essential at relatively small energies of neutral pions.  To suppress
the relative part $u$ (close to one) of such a background, we demand
that the cone with a specially defined total angle $\theta(u)$ around
the photon propagation should cross the detector plane inside of it.
The angle $\theta(u)$ can be easily estimated from the isotropy of
$\pi^0\to\gamma\gamma$ decay in the meson rest frame after the Lorentz
boost to the laboratory system
\begin{equation}
  \theta(u)=\arctan\left(\frac{m_{\pi^0}(1+\beta u)\sqrt{1-u^2}}
                              {2E(\beta+u)}\right)+
            \arctan\left(\frac{m_{\pi^0}(1+\beta u)\sqrt{1-u^2}}
                              {2E(\beta-u)}\right),
  \label{theta_cone}
\end{equation}
where $E$ is the energy of the analyzed cluster, $\beta\simeq 1$ is
the $\pi^0$-meson velocity in units of the light speed.  Note that
minimal decay cone angle $\theta_{\min}=2\arccos (\beta)$ can be
obtained from formula (\ref{theta_cone}) at $u=0$.  In the case when
both photons fly towards the detector direction we arrive at
$u<\beta$.  To apply the considered cut we take $u=0.95$.  This cut
actually works at small energies, and mainly at the edges of the
calorimeter, while at high energies practically all events pass it
successfully.
 
The second background comes from the limited spatial resolution of
PHOS, i.e. when the overlapping showers from two-photon decays of the
high-energy pions form one cluster.  This mechanism provides the major
background for direct photons at high energies. Its suppression is
mainly fulfilled by the ANN-algorithm.  The step sequence runs as
follows.

First we apply the procedure of the local peak number determination in
a cluster. It classifies the cluster as that with two overlapping
showers when two local peaks are observed above the electronic noise
fluctuations.  Such classification of clusters is a common algorithm
in reconstruction programs for cellular detectors.

To decrease the background from the decays $\pi^0\to\gamma\gamma$,
when only one photon hits PHOS due to the limited aperture, we apply
cut (\ref{theta_cone}) with the use of $\theta(u)$.  The remaining
clusters with one peak passing this 2-stage preselection are analyzed
by ANN.  It was composed of three layers (see Fig.~\ref{fig:ANN}):
\begin{figure}[htbp]
  \parbox{0.5\hsize}{
    \includegraphics[width=\hsize]{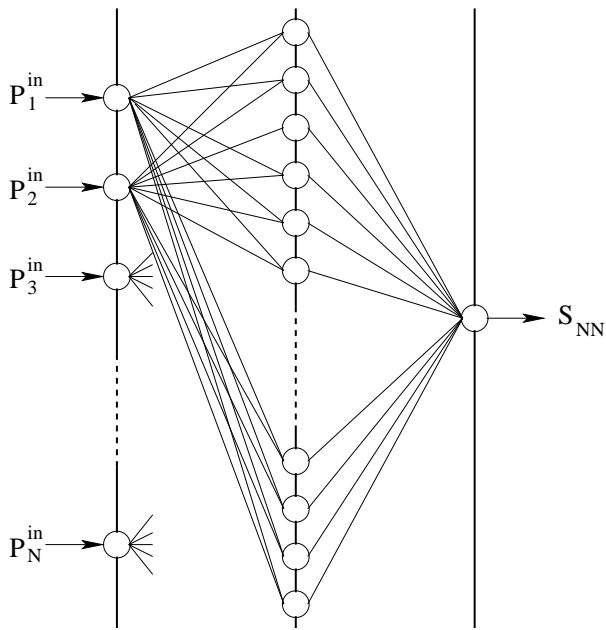}}
  \hfill
  \parbox{0.45\hsize}{
    \caption{Architecture of the used ANN.}
    \label{fig:ANN}
  }
\end{figure}
input, hidden and output. The input layer consists of $N$ nodes, where
$N$ is the dimension of vector $\vec{P}^{(in)}$ representing the event
features; the hidden layer is built of $2N+1$ nodes; and finally, the
one-node output layer provides the neural-net response $S_{NN} \in
(0,1)$.

The net was trained with two samples of events containing the desired
signals (i.e. clusters from direct photons) and the background (i.e.
clusters produced by the overlapped photons from decay $\pi^0
\rightarrow 2 \gamma$).  Each sample consisted of 30{,}000 events
(clusters) left after preselection.  Upon training the net, we tested
its efficiency on another statistically independent signal and
background samples, each consisting of 30{,}000 events also left after
the preselection.

\section{Simulation of isolating photons and $\pi^0$-mesons}

The data taken for the analysis were simulated within the ALICE
framework for the simulation, reconstruction and data analysis, {\tt
  aliroot v.3.06.02} \cite{aliroot}.  Two samples of events were
generated, one sample containing one photon and another sample having
one $\pi^0$ per event.  Photons and $\pi^0$'s were emitted from the
interaction point with the uniformly distributed transverse momentum
in the range of $0 < p_T < 120$~GeV/c within the solid angle defined
by the uniformly distributed azimuth angle $210^\circ < \phi <
330^\circ$ and the polar angle $80^\circ < \theta < 100^\circ$. The
solid angle of the emitted particles was chosen to be a little larger
that that of PHOS detector, to avoid various border effects. The decay
of $\pi^0$'s was performed by the {\tt aliroot}.

\begin{figure}[htb]
  \centerline{
    \includegraphics[width=0.8\hsize]{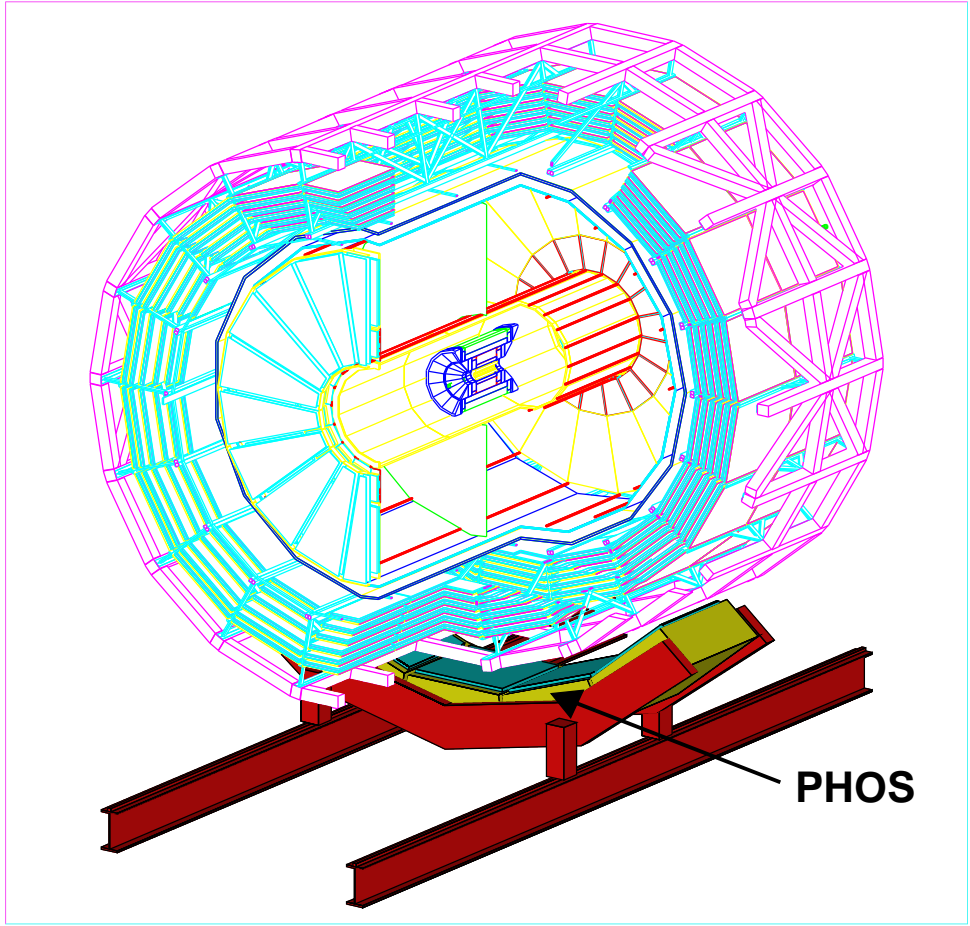}
  }
  \caption{ALICE detector used for the simulation in {\tt aliroot}.}
  \label{fig:ALICE}
\end{figure}

The real ALICE environment was taken into account during the particle
tracking from the interaction point to PHOS. The following
detectors and infrastructure modules which cover the PHOS aperture
were installed: PIPE, ITS, TPC, TRD, FRAME, as shown in
Fig.\ref{fig:ALICE}.
This environment results in particle interactions with the media and
the production of secondary particles which deteriorate the response
of the PHOS detector.

The response of the PHOS calorimeter was simulated by GEANT 3.21
\cite{GEANT321}, which was included into the {\tt aliroot} package as
a particle tracking and physics interaction tool.  The showers
developed by the particles passing through the calorimeter cells, gave
the {\tt HITS} which were the deposited energy of the shower in each
cell. These {\tt HITS} were digitized, i.e. the energy amplitude of
the cell was measured by the two-channel ADC, and the integer signal
of the ADC produced {\tt DIGITS}. The noise with $\sigma_{\rm
  noise}=10$ MeV was applied to the {\tt DIGITS}, after which the {\tt
  DIGITS} were passed through the 50-MeV threshold cut.  The remaining
{\tt DIGITS} with a common side were grouped into clusters. Only
clusters with the total energy greater than 500 MeV were accepted.

The data that passed to the Neural Network analysis contained the
following information. The event samples (photons or $\pi^0$'s) were
identified by the file name. Events had a header which was
characterized by the event number, the energy of the produced particle
and the number of the found clusters. The event header was followed by
the cluster data consisting of the cluster header with the cluster
number and the number of cells in the cluster, and the cell
information containing the list of cell positions in PHOS as well as
the cell amplitudes.

\section{Results of the $\gamma/\pi^0$-selection}

Analysis of the generated data shows that only 0.04\% of direct
photons are reconstructed as two-cluster events. The energies of the
additional clusters are limited, they are practically not greater than
1.5 GeV.

\begin{figure}[htbp]
  \centerline{
    \includegraphics[width=0.85\hsize,bb=40 20 525 520]{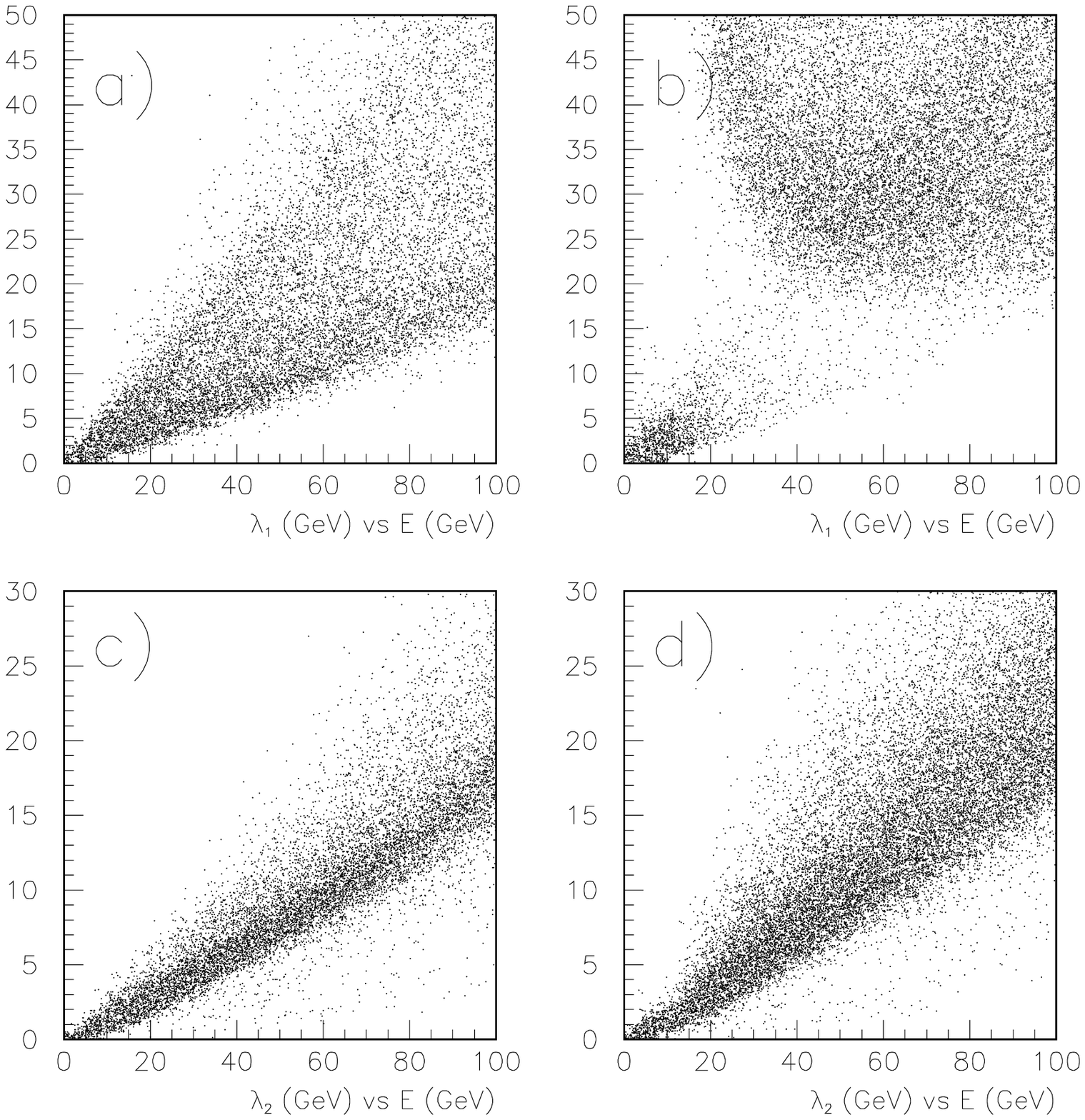}}
  \caption{Two-dimensional plots of $\lambda_1$ (a), (b) and 
    $\lambda_2$ (c), (d) $vs$ energy, (a) and (c) are direct photons, 
    (b) and (d) are the background. 
  }
  \label{l1l2vse}
%\end{figure}
%
\medskip
%\begin{figure}[htbp]
  \centerline{
    \epsfxsize=0.85\hsize \epsfbox[30 270 520 520]{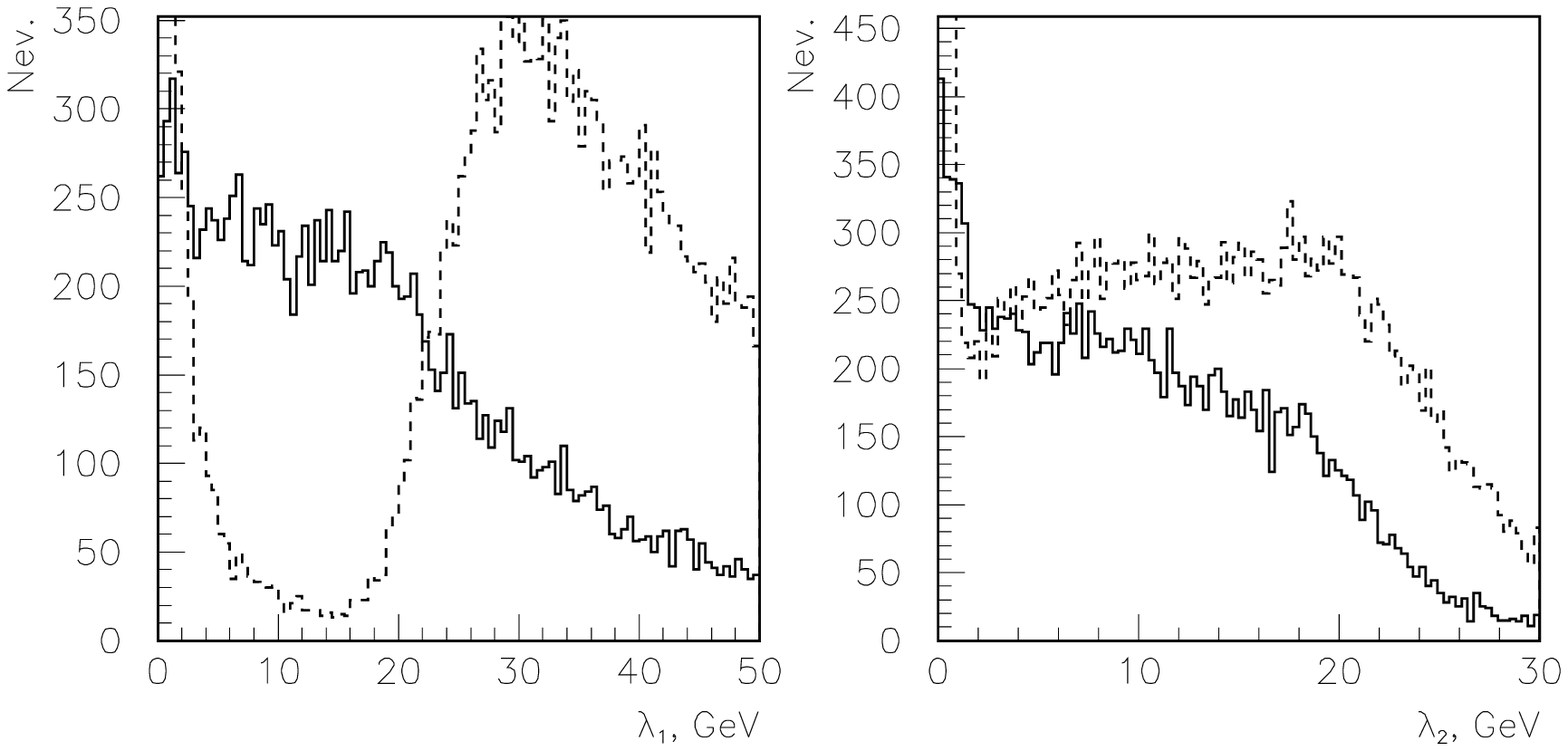}
  } 
  \caption{$\lambda_1$- and $\lambda_2$ distributions for
    direct photons (solid lines) and background (dotted lines).}
  \label{l1l2h}
\end{figure}
\begin{figure}[htb]
  \centerline{
    \epsfxsize=0.9\hsize 
    \epsfysize=13cm
    \epsfbox[0 15 550 790]{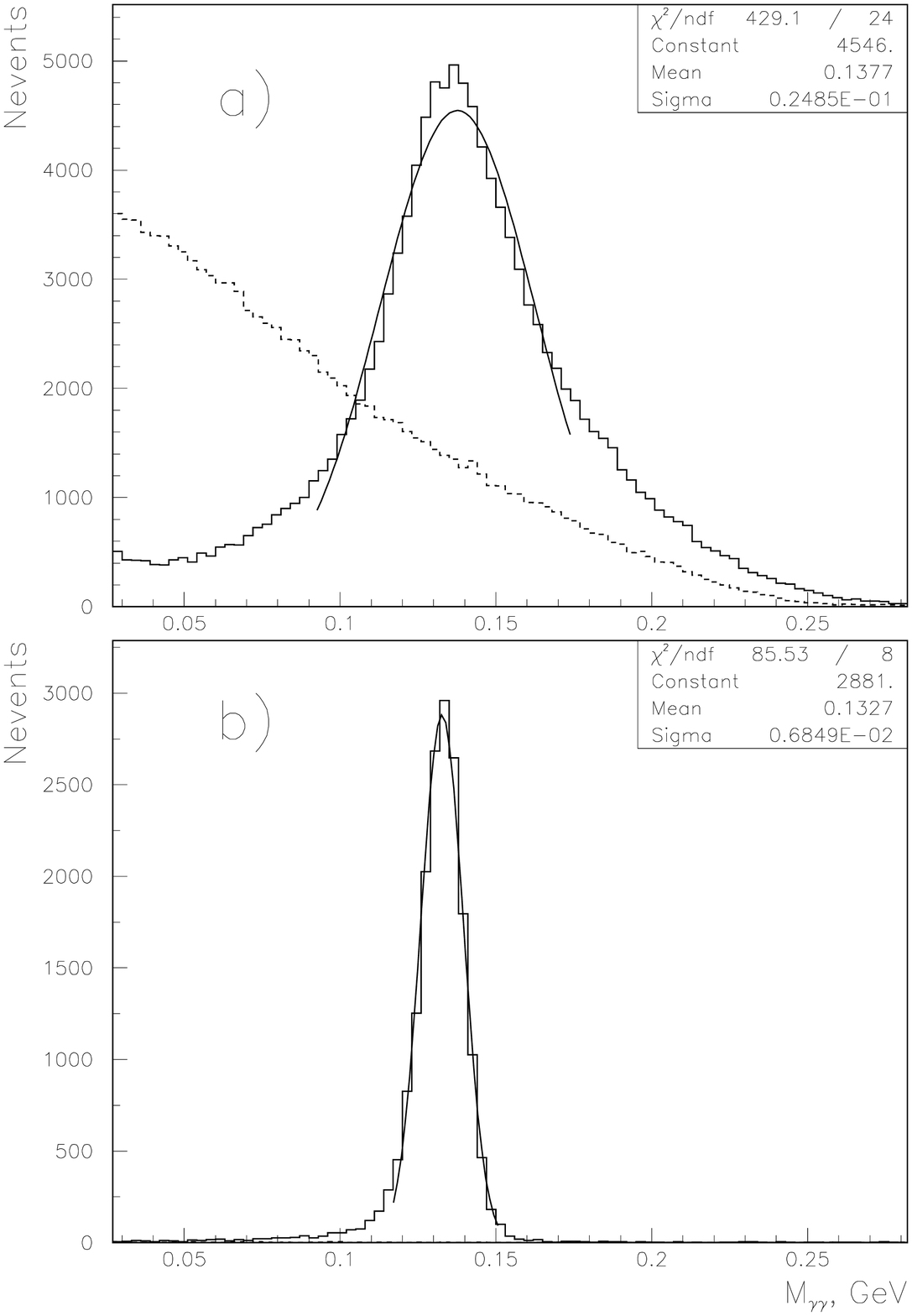}
  }
  \caption{ a) Distribution illustrating the  
    results of calculating $\gamma\gamma$-effective masses with
    formula (\protect{\ref{mef_for}}) for overlapped clusters from the
    neutral pion decays (solid lines) and the same data for the sample
    of direct photons (dotted lines).  b) Distribution of
    $\gamma\gamma$-effective masses for the divided clusters from
    $\pi_0$ decays. A mistaken contribution from the prompt photon due
    to the splitting of clusters under treatment is negligible on the
    level of unity events in the $\pi_0$-region.  Smooth solid curves
    show the Gaussian fit with the average value of 133 MeV and the
    variation of 6.8 MeV.}
  \label{mef_fig}
\end{figure}

Two-dimensional plots (Fig.\ref{l1l2vse}) of eigen values $\lambda_1$
and $\lambda_2$ versus energy $E$ of $\gamma$ or $\pi^0$, demonstrate
the difference between the direct photons and the background that is
used in the process of the ANN training.  Fig. \ref{l1l2h} also shows
one-dimensional distributions which demonstrate a rather
well-expressed $\gamma/\pi^0$-distinction.

Fig.\ref{mef_fig} shows comparative results of the calculations of
$\gamma\gamma$-effective masses $M_{\gamma\gamma}$ for overlapped and
separated clusters from $\pi^0$-decays and the same data from the
sample of direct photons. For the overlapped clusters we use formula
(\ref{mef_for}) expressing $M_{\gamma\gamma}$ through the moments
$M_{mn}$. In both cases for the overlapped and separated clusters from
$\pi^0$-decays there are well observed peaks at the value of
$\pi_0$-mass while application of the same formula (\ref{mef_for}) for
direct photons gives a smoothly falling dependence with the growth of
mass in the region of the $\pi_0$-meson.  Smooth solid curves in the
figure show the results of Gaussian fits of $\pi^0$-peaks. The average
value and variation for the separated clusters are equal to 133 MeV
and 6.8 MeV, respectively.
 
\begin{figure}[htbp]
  \centerline{
    \includegraphics[width=0.9\hsize]{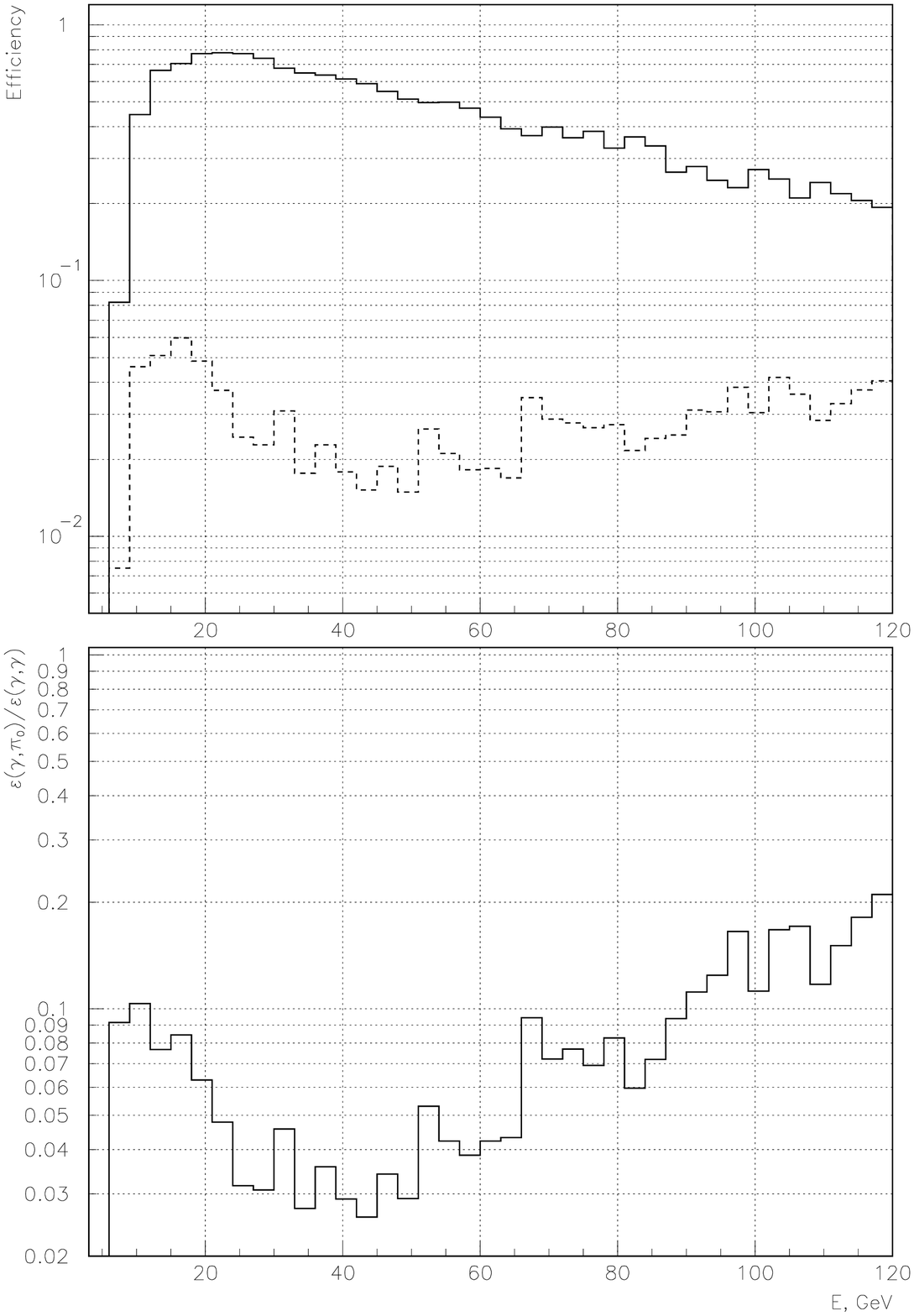}
  }
  \caption{
    Efficiency $\varepsilon (\gamma,\gamma )$ of true photon
    identification as a photon (solid lines), misidentification
    $\varepsilon (\gamma,\pi ^0)$ of $\pi^0$-meson as a photon (dotted
    lines), and the coefficient of the background suppression
    $\varepsilon (\gamma,\pi ^0)/\varepsilon (\gamma,\gamma )$ as a
    functions of $E$.  The vector of event features is $(E, \lambda_1,
    \lambda_2, M_{30}, M_{04}, \phi )$.  }
\label{eff1}
\end{figure}

The main results of this article concerning the quality of the ANN
training for the $\gamma/\pi^0$-discrimination are presented in Fig.
\ref{eff1} for one of the possible feature event vectors, selected as
$(E, \lambda_1, \lambda_2, M_{30}, M_{04}, \phi)$, where $E$ is the
measured cluster energy.  The cut on the $S_{NN}$ signal equal to
0.64.  This figure shows the efficiency $\varepsilon(\gamma,\gamma)$
of true photon identification as a photon, misidentification
$\varepsilon(\gamma,\pi^0)$ of $\pi^0$-meson as a photon in the range
of generated energies of photons and $\pi^0$-mesons from 3 GeV to
120~GeV, and the coefficient of background suppression relatively to
direct photons $\varepsilon (\gamma,\pi^0)/\varepsilon
(\gamma,\gamma)$.
One can see that the probability $\varepsilon (\gamma,\pi^0)$ of
misidentification of a neutral pion as a photon is on the level of a
few percent in the energy range of $3-120$~GeV with relatively high
efficiency of the true photon identification.  The rise of the $\pi^0$
misidentification probability at energies below 25~GeV is caused by
the $\pi^0 \to \gamma\gamma$ decays with one photon outside PHOS,
whereas the decrease of the true photon identification efficiency at
the energies below $15$~GeV is the result of the true photon cut
decreasing the background from these decays of $\pi^0$'s with only one
photon inside PHOS.
 
We compare our results with the data of work \cite{Minaev} where the
coefficient $\varepsilon(\gamma,\pi^0)$ was estimated for the STAR
experiment as 0.15 at 20 GeV and 0.45 at 40 GeV at fixed
$\varepsilon(\gamma,\gamma) = 0.8$.  The efficiency of the
$\gamma/\pi^0$-recognition was also calculated within the neural
network approach for the CMS experiment in note \cite{CMS_NN}.  The
obtained values $\varepsilon(\gamma,\pi^0)$ varied from 0.25 to 0.55
at $E =20$~GeV and from 0.40 to 0.55 at $E =100$ GeV, dependent on the
rapidity range, while the efficiency of the single photon recognition
was kept at 91\%.
   
\clearpage
\section{Conclusion}
In this paper a neural network method is developed to separate the
direct photons from the neutral pion background in the PHOS
spectrometer of the ALICE experiment.  The proposed algorithm is based
on the analysis of the energy-profile tensor of the cluster calculated
in its eigen vector coordinate system.  The proposed method allows to
construct effective event feature vectors consisting of a limited
number of variables carrying enough information to train the neural
network for the goals of $\gamma/\pi^0$ separation.  This method has
been applied for Monte-Carlo events in PHOS.  It has been found that
the probability of misidentification of a neutral pion as a photon is
on the level of a few percent in the pion energy range of $3-120$~GeV
with the relatively high efficiency of the correct photon
identification as an isolated photon in the same energy range.

\clearpage
\section*{Addendum.\\
  \medskip Calculation of $\gamma\gamma$-effective mass through
  moments $M_{mn}$} \setcounter{equation}{0}

\renewcommand\theequation{A.\arabic{equation}}

Kinematics formulae can be applied to estimate the
$\gamma\gamma$-effective mass $M_{\gamma\gamma}$ in the case of the
divided clusters on the base of the measured energies of photons and
the angle between their momenta. The problem is more complicated for
the overlapped clusters.  Here we follow paper \cite{Bituyukov} to
express the mass $M_{\gamma\gamma}$ in terms of the above introduced
moments (\ref{M}) under assumption that the shower profile for an
isolated photon possesses an azimuthal symmetry with respect to the
point of $\gamma$-hit. This leads to the following properties of the
own moments $\bar M_{mn}$ (i.e. moments relative to the point of
$\gamma$-hit) for the isolated showers: $\bar M_{m0}=\bar M_{0m}$ and
$\bar M_{mk}=0$ for odd $k$.  The $\gamma$-incidence inclination
violates the azimuthal symmetry, but we restore it by the mentioned
compression of the cluster space (\ref{compression}).

Due to the principle of energy additivity in the case of the
overlapped showers the summary cluster energy density
$F(E_1,E_2,x_1,x_2)$ can be expressed in the eigen vector coordinate
system of the cluster as follows:
\begin{equation}
  F(E_1,E_2,x_1,x_2)=E_1f(x_1-x_{10},x_2)+E_2f(x_1-x_{20},x_2)~,
  \label{Fff}
\end{equation}
where $E_1$, $E_2$ are the individual energies of showers, $x_1$ and
$x_2$ are the coordinates of an arbitrary point in the cluster space,
$x_{10}$ and $x_{20}$ are the coordinates of photon hits along the
eigen vector $\vec{e}_1$, and $f(x_1,x_2)$ is the single photon shower
profile. After that one can easily obtain from (\ref{M})
\begin{equation}
  M_{ln}=\sum\limits_{i=0}^{l} C_l^i~
  [~E_1^{}x_{10}^{l-i}\bar m_{in}+E_2^{}x_{20}^{l-i}\bar m_{in}~]~,    
  \label{Msum}
\end{equation} 
where a non-zero contribution gives only terms with even $i$;
$C_l^i=l!/\left(i!(l-i)!\right)$ are binomial coefficients, $\bar
m_{in}$ are normalized own moments $\bar m_{in} =\bar M_{in}/M_{00}$
of a single photon shower.  Taking into account the above mentioned
azimuthal symmetry of single photon showers, we obtain the following
set of equations for the determination of energies and coordinates of
individual showers $E_1$, $E_2$, $x_{10}$, $x_{20}$:
\begin{equation}
  \cases{
    E_1~+~E_2~=M_{00} \cr
    E_1\cdot x_{10}+E_2\cdot x_{20}~=M_{10}\cr
    E_1\cdot x_{10}^2+E_2\cdot x_{20}^2 =M_{20}-M_{02}\cr
    %E_1\cdot x_{10}^3+E_2\cdot x_{20}^3=M_{30}-M_{02}M_{10}/M_{00}\cr
    E_1\cdot x_{10}^3+E_2\cdot x_{20}^3=M_{30}-3M_{12}
  }
  \label{M20}
\end{equation}
The solution is: 
\begin{equation}
  \cases{      
    \displaystyle
    X=\frac{M_{30}-3M_{12}-M_{10}(M_{20}-M_{02})/M_{00}}
    {M_{20}-M_{02}-M^2_{10}/M_{00}}\cr
    d^2=X^2-4(XM_{10}-M_{20}+M_{02})/M_{00}\cr
    \Delta=(2M_{10}-XM_{00})/d\cr
    E=M_{00},\cr
  }
  \label{solution1}
\end{equation} 
where $X=x_{20}+x_{10}$, $d=x_{20}-x_{10}$, $\Delta=E_2-E_1$ and
$E=E_2+E_1$.  The variable $d$ gives the estimation of distance
between the hits of individual showers in the overlapped cluster.  To
calculate the effective $\gamma\gamma$-mass one should also know the
distance $R$ from the interaction point to the cluster center to
determine the decay angle $\theta$ between $\gamma$-quanta, and thus
we obtain
\begin{equation}    
  M_{\gamma\gamma}^2=4E_1E_2\sin^2(\theta/2)=(E^2-\Delta^2)
  \frac{d^2}{d^2+4R^2}.
  \label{mef_for}
\end{equation} 
        
\end{document}